\documentclass[5p,times,twocolumn]{elsarticle}        

\usepackage{preamble_paper}
\usepackage{newcommands_paper}

\begin{document}

\title{Zinc Electrode Shape-Change in Secondary Air Batteries: A 2D Modeling Approach}

\author[dlr,hiu,uu]{Tobias Schmitt}
\author[tub]{Tobias Arlt}
\author[hzb]{Ingo Manke}
\author[dlr,hiu,uu]{Arnulf Latz}
\author[dlr,hiu,uu]{Birger Horstmann\corref{cor1}}\ead[url]{birger.horstmann@dlr.de}

\cortext[cor1]{Corresponding author}
\address[dlr]{German Aerospace Center, Pfaffenwaldring 38-40, 70569 Stuttgart, Germany}
\address[hiu]{Helmholtz Institute Ulm, Helmholtzstra\ss{}e 11, 89081 Ulm, Germany} 
\address[uu]{Ulm University, Institute of Electrochemistry, Albert-Einstein-Allee 47, 89069 Ulm, Germany}
\address[tub]{Technical University of Berlin, Material Sciences and Technology, Hardenbergstra\ss{}e 36, 10623 Berlin, Germany}
\address[hzb]{Helmholtz Center for Materials and Energy, Hahn-Meitner-Platz 1, 14109 Berlin, Germany}

\begin{abstract}
Zinc-air batteries offer large specific energy densities, while relying on abundant and non-toxic materials. In this paper, we present the first multi-dimensional simulations of zinc-air batteries. We refine our existing theory-based model of secondary zinc-air systems. The model comprises thermodynamically consistent multi-species transport in alkaline electrolytes, formation and dissolution of metallic zinc and passivating zinc oxide, as well as multi-phase coexistence in gas diffusion electrodes. For the first time, we simulate zinc shape-change during battery cycling by modeling convection of zinc solids. We validate our model with in-situ tomography of commercial button cells. Two-dimensional volume-averaged simulations of cell voltage and zinc electrode morphology during discharge agree with these measurements. Thus, we can study how electrolyte carbonation limits shelf-life and how zinc shape-change limits cycle-life. The charging current is found to be the major contributor to cycle-life limitations. Finally, we optimize initial anode structure and charge-discharge protocols for improved performance and cycle-ability.
\end{abstract}


\maketitle

\section{Introduction}
\label{sec:intro}
Electrochemical energy storage plays a key role for the transition to a carbon neutral society \cite{Groger2015}. Energy-efficient lithium-ion batteries are optimally suited for portable electronics and local electric mobility. Scalable electrolysers and fuel cells address the needs of long-term grid-scale energy storage. In between, metal-air batteries promise high specific energy densities at low costs for household-level energy storage or range extenders \cite{Li2014,Li2017,Clark2018}. Possible material choices are, with increasing energy density, iron, zinc, magnesium, aluminum, and lithium. A couple of years ago, research concentrated on lithium-air batteries because of its high theoretical energy densities \cite{grande2015lithium,horstmann2013rate,Horstmann2013}. Unfortunately, fundamental challenges still prevent their commercialization. The reasons include low electrolyte stability under oxygen evolution conditions, slow growth and dissolution of lithium peroxide, low oxygen solubility in the electrolytes, and shape-change of lithium metal.

Zinc-air batteries are a promising alternative. Zinc is cheap, ecologically friendly, nonflammable, globally abundant, and non-toxic \cite{Li2014,Li2017,Yi2018}. Most importantly, zinc metal is stable in aqueous electrolytes aside from controllable hydrogen evolution \cite{lysgaard2018combined}. For aqueous electrolytes, gas diffusion cathodes offer fast oxygen transport \cite{Horstmann2013,Danner2016}. Therefore, primary zinc-air cells are commercially available for decades. However, their cycle-life and shelf-life is limited. Firstly, the standard alkaline electrolytes absorb carbon dioxide from air, which reduces their pH value and limits shelf-life to a few months \cite{Stamm2017}. Possible mitigation methods include carbon dioxide absorbers \cite{Drillet2001} and novel electrolytes \cite{mainar2018overview}, e.g., near-neutral ammonium chloride solutions \cite{Goh2014,Sumboja2016,Clark2017,Clark2019} or ionic liquids \cite{ingale2017test,Hoffmann2018,Wang2018}. Secondly, porous zinc anodes alter their shape during battery cycling, which reduces porosity and limits cycle-life \cite{Yi2018}. Novel zinc structures are proposed to reduce shape-change \cite{mainar2018brief}, e.g., hyper-dendritic zinc foam \cite{Chamoun2015}, 3D zinc sponge \cite{parker2017rechargeable}, or mechanical coatings \cite{Stock2018,Michlik2018}.  

In this article, we present the first multi-dimensional simulations of zinc-air batteries. Our 2D simulations of button cells are enabled by local volume-averaging theory \cite{Chen1995,Whitaker1999}. We optimize cell design and complement experimental efforts. Our electrochemical model elucidates the intimate correlation between electrode morphology and electrolyte dynamics during battery cycling. We parametrize and validate our model by x-ray tomography data \cite{Arlt2014,Schroder2014_2}. Recently, we developed a theory-based model of aqueous alkaline zinc-air batteries that consistently couples the dynamics of its solid, liquid, and gaseous phases \cite{Stamm2017}. 1D simulations predicted the effects of inhomogeneous zinc oxide nucleation and growth as well as carbonate formation in the electrolyte. We validated the model predictions of discharge voltage and shelf-life. Here, we extend our model by convection of solid zinc metal and simulate zinc shape-change in secondary zinc-air batteries.

Preceding zinc-air battery models \cite{Schroder2014_2,Sunu1980,Isaacson1990, Mao1992,Deiss2002,Schroder2014_1} did not achieve such a level of consistency and predictive powers as discussed in Refs. \cite{Clark2018,Stamm2017}. These 0D or 1D simulations neglect electrochemical interactions and local material inhomogeneities. However, 3D simulations advance in the field of lithium-ion batteries \cite{Latz2011,Latz2015,Bauer2011,cogswell2013theory,Hein2016_1,Hein2016_2,finsterbusch2018high}. A few multi-dimensional models are applied to conversion type batteries, mainly analyzing single electrodes. For lithium-air batteries, our group performs 3D Lattice-Boltzmann simulations of phase coexistence in gas diffusion electrodes \cite{Danner2016}. Dendrite formation in lithium metal anodes is measured and simulated in Ref. \cite{Biton2017}. We are aware of a single 3D simulation of a lithium-air battery cell, which has a limited resolution and shows only slight 3D effects \cite{Gwak2016}. 

We proceed with descriptions of our x-ray experiments, our theory-based model, and our computational implementation in Sec. \ref{sec:es}, Sec. \ref{sec:mod}, and Sec. \ref{sec:cd}, respectively. We simulate discharge at various currents for comparison with tomographies \cite{Arlt2014} and 1D simulations \cite{Stamm2017} (see Sec. \ref{sec:rse}). Finally, we discuss strategies to reduce zinc shape-change in Sec. \ref{sec:do}.

\section{Experiment}
\label{sec:es}
\begin{figure}[t]
	\centering
			\includegraphics[width = 0.97\columnwidth]{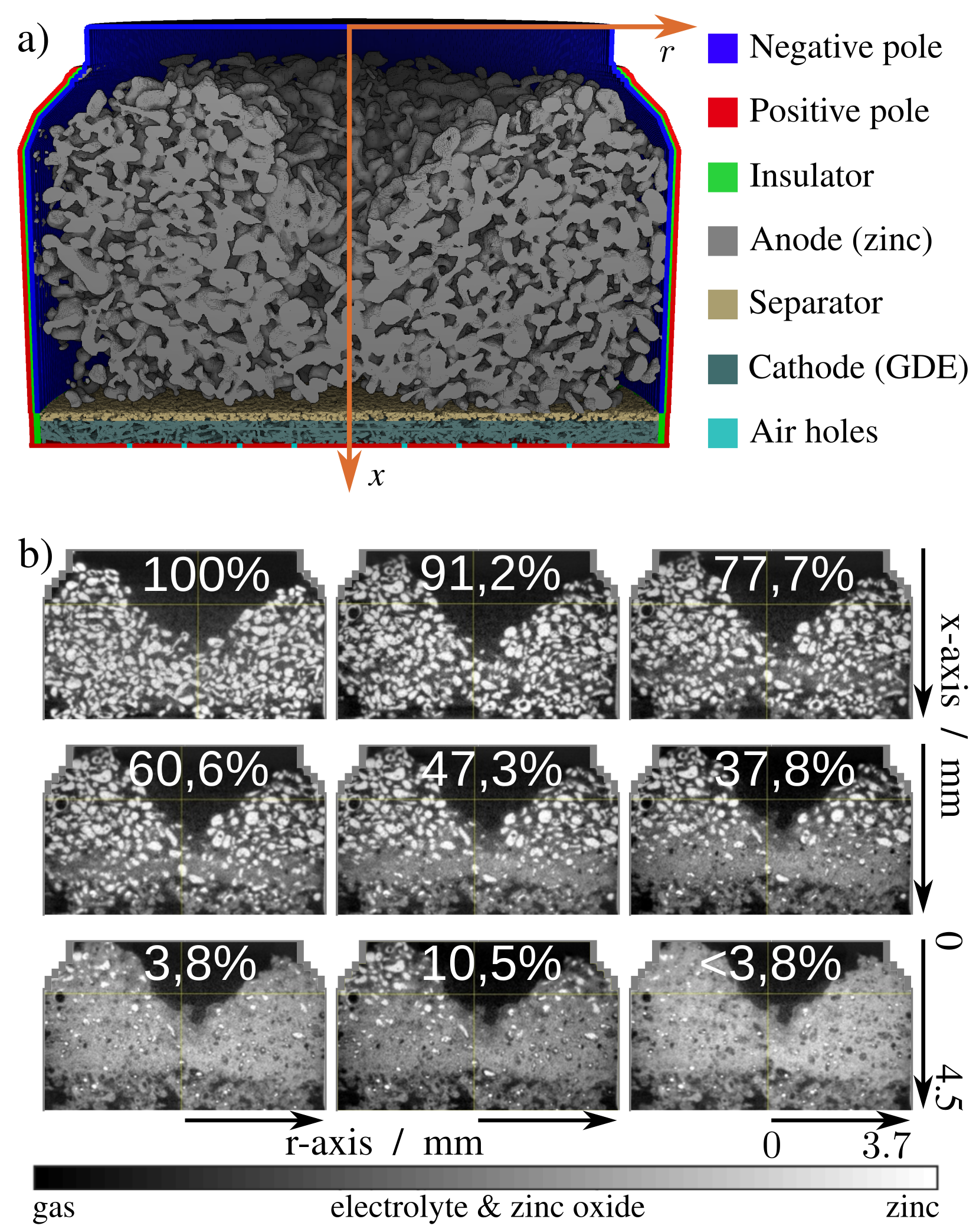}
	\caption[Cross-sections .]{(a) Schematic image of a fully charged zinc-air button cell. (b) Cross-sections of the reconstructed tomographies for a zinc electrode of a Renata ZA13 cell during discharge with current $\myvec{j} = \SI{54.2}{\ampere \per \meter\squared}$.}
	\label{fig:es:tomography}
\end{figure}

We simulate the 
Renata ZA13 button cell \cite{RENATA_1} (see \cref{fig:es:tomography}a). Air flows through the lids into the gas diffusion electrode (GDE), where it dissolves in the electrolyte and the oxygen reaction is catalyzed. Most of the cell volume is occupied by the negative electrode. In the fully charged state, it contains a porous zinc powder. The void space in the top region below the negative pole compensates the volume expansion from zinc to zinc oxide during discharge.

For parametrization and validation, we discharge the Renata battery \cite{RENATA_1} using a Gamry 3000 potentiostat at an ambient temperature of $T=20$\dg C. The galvanostatic discharge current per cross section area is \SI{54.2}{\ampere \per \squared\meter}. The cross section of the cell on the cathode side is $\sim\SI{4.2e-5}{\meter \squared}$. A conventional X-ray tube is used for tomographic measurements in combination with a Hamamatsu X-ray flat panel detector with $2316 \times 2316$ pixel, resulting in a pixel size of \SI{9}{\micro \meter}. Details of the setup are given in \cref*{fig:suppl:misc:setup} and Refs. \cite{Arlt2014,Schroder2014_2,Franke-Lang2017}. An accelerating voltage of \SI{120}{\kilo \volt} is applied while the current at the tungsten anode is adjusted to \SI{84}{\micro \ampere}. A \SI{0.5}{\milli \meter} Cu filter is used in order to avoid beam hardening. We take 1200 projections for a full three-dimensional tomography. Each angle step is exposed for three times for \SI{1}{\second}  in order to increase the signal-to-noise-ratio. 

We perform tomographic imaging at various state-of-charges during discharge until the cut-off potential \SI{0.6}{\volt} is reached (see \cref{fig:es:tomography}b). Before each tomography, the applied current is switched off and the voltage is allowed to relax. Each tomography under steady-state condition takes about \SI{1.5}{\hour}. We describe data processing in \cref{sec:cd:init} below.

\section{Theory and Model}
\label{sec:mod}
In this section, we present a thermodynamically consistent and volume-averaged multi-dimensional model for rechargeable zinc-air cells. It is based on our 1D model developed in Refs. \cite{Stamm2017,Horstmann2013}. We model the phase coexistence of solid ($\alpha=\sol$), liquid ($\alpha=\liq$), and gaseous phases ($\alpha=\gas$). For each state of matter $\alpha$, we consider distinct species $\beta_{\alpha}$. 

As solids, we take into account zinc ($\beta_{\sol} = \zn$), zinc oxide ($\beta_{\sol} = \zno$), and  passive material ($\beta_{\sol} =\passive$), e.g., housing, separator. Our recent 1D model discussed nucleation, growth, and dissolution of zinc and zinc oxide \cite{Stamm2017}. We simulated how zinc oxide passivates zinc particles. In this article, we present the first model of zinc convection and simulate shape-change. 

The electrolyte is a concentrated aqueous potassium hydroxide solution. Its constituents are water ($\bl=\water$), potassium ($\bl=\kalium$), hydroxide ($\bl=\oh$), zincate ($\bl=\znoh$), carbonate ($\bl=\co$), and dissolved oxygen ($\bl=\oli$). We denote the neutral, positive, and negative species with the indices $\bln$, $\blp$, $\blm$, respectively. Our theory-based modeling reveals that local charge and volume constraints reduce the number of independent species \cite{Stamm2017}. We choose water and potassium as dependent species and reserve the index $\gamma_{\alpha}$ for the independent species. A multi-component volume-constraint induces electrolyte convection.

From the gaseous phase, we model the inflow of oxygen ($\beta_{\gas} = \og$) and carbon dioxide ($\beta_{\gas}={\cotwo}$). The gaseous phase makes possible fast oxygen transport through the GDE \cite{Horstmann2013}. The electrolyte degrades because of carbon dioxide absorption. A gas reservoir on top of the zinc anode buffers the volume expansion from zinc to zinc oxide.

In the following, we present our electrochemical transport and reaction model based on Ref. \cite{Stamm2017}. In \cref{sec:mod:lva}, we discuss the volume-averaging process that makes possible our multi-dimensional simulations. We derive equations for the transport of the solid phases ($\zn$, $\zno$, $\passive$) in \cref{sec:mod:lva:sol}, of the liquid phase in \cref{sec:mod:lva:liq}, and of the gas phase in \cref{sec:mod:lva:gas}. Finally, we model the relevant reactions in \cref{sec:mod:reac}.

\subsection{Volume-Averaged Model}
\label{sec:mod:lva}
Because on the micro-scale of a zinc-air cell multiple distinct phases are entangled, simulating their dynamics is a challenging task. We develop efficient multi-dimensional simulations based on the approach of volume-averaging as discussed by Whitaker and Chen \cite{Chen1995,Whitaker1999,Whitaker1984}. Thus, we treat the composite systems as a bulk of mixed materials. In this article, we present a simplified derivation of local volume-averaging and refer to an upcoming publication for further details. 

Our transport theory relies on microscopic single-phase dynamics. In the Euler frame, i.e., lab frame, the continuity equation holds
\begin{align}
\label{eq:mod:tt:global_conservation_equation_euler_loc} 
&\partial_{t} \Psi = - \bn \left( \ncflux + \Psi \vel \right)  \, , 
\end{align}
where $\Psi(t,x)$ represents the conserved variables, e.g., concentration. $\vel$ is the center-of-mass convection velocity and $\ncflux$ is the non-convective flux density. Here, we omit source terms $\source^{\Omega}=0$, as we get by without homogeneous bulk reactions. 

We choose an averaging domain $V$ which exceeds the typical size of the micro-structure. Firstly, we define the superficial volume-average of a field variable $\Psi_{\alpha}$ in phase $\alpha$ as average over the composite domain
\begin{align}
\lvaSup{\Psi_{\alpha}} &\equiv \lvaSup{\Psi_{\alpha}} (t,\myvec{x}) = \frac{1}{\abs{V}} \int\limits_{V_{\alpha}} \Psi_{\alpha}(t,\myvec{x} - \myvec{x}') d^{3}\myvec{x}' \, ,
\label{eq:mod:lva:lvaSup}
\end{align}
where the integral is restricted to single phases $V_{\alpha}$ within the domain $V$. The single-phase continuity equation \cref{eq:mod:tt:global_conservation_equation_euler_loc}, valid in phase $\alpha$, transforms into the volume-averaged form
\begin{align}
\label{eq:mod:lva:con}
\partial_{t} \lvaSup{\Psi_{\alpha}} = -\bn\cdot \lvaSup{\ncflux_{ \Psi_\alpha}} - \bn\cdot \lvaSup{\Psi_{\alpha}\vel_\alpha} + \lvaSurfC{\source}^{\Theta}_{\Psi_\alpha}  \, ,
\end{align}
with the surface reaction rates $\lvaSurfC{\source}^{\Theta}_{\alpha}$ specified in \cref{sec:mod:reac}. In the next three sections, \cref{sec:mod:lva:sol}, \cref{sec:mod:lva:liq}, and \cref{sec:mod:lva:gas}, we express the superficial volume-averages of the transport terms with averages over single phases.  

Secondly, we define the intrinsic volume-average $\lvaIn{\Psi}_{\alpha}$ and the local surface-average $\lvaSurfC{\Psi}_{\alpha}$ as average over a single phase
\begin{align}
	\lvaIn{\Psi}_{\alpha} &\equiv \lvaIn{\Psi}_{\alpha} (t,\myvec{x}) = \frac{1}{\abs{V_{\alpha}}} \int\limits_{V_{\alpha}} \Psi_{\alpha}(t,\myvec{x} - \myvec{x}') d^{3}\myvec{x}' \, ,
	\label{eq:mod:lva:lvaIn}\\
	~
	\lvaSurfC{\Psi}_{\alpha} &\equiv \lvaSurfC{\Psi}_{\alpha} (t,\myvec{x}) = \frac{1}{\abs{A_{\alpha}}} \int\limits_{A_{\alpha}} \Psi_{\alpha}(t,\myvec{x} - \myvec{x}') d^{2}\myvec{x}' \, ,
	\label{eq:mod:lva:lvaSurfC}
\end{align}
with the surface $A_{\alpha}$ of phase $\alpha$ within the domain $V$. The two volume-averages are connected by the volume fractions $\varepsilon_\alpha$
\begin{align}
\lvaSup{\Psi_{\alpha}} = \frac{\abs{V_{\alpha}}}{\abs{V}} \lvaIn{\Psi}_{\alpha} = \varepsilon_{\alpha} \lvaIn{\Psi}_{\alpha}  \, .
\label{eq:mod:lva:connLVA_sup_intr}
\end{align}
The volume fractions sum to unity
\begin{align}
\lvaSup{1} = \sum\limits_{\alpha} \lvaSup{1_{\alpha}} = \sum\limits_{\alpha} \varepsilon_{\alpha} = \epsL + \varepsilon_{\zn} + \varepsilon_{\zno} + \varepsilon_{\passive} + \epsG  \, .  \label{eq:mod:lva:sum_volume_fraction}
\end{align}
In the following, we discuss the transport equations for solid, liquid, and gaseous phases.

\subsection{Solid Phase Transport}
\label{sec:mod:lva:sol}

\begin{figure}[!t]
\centering
	\includegraphics[width=\columnwidth]{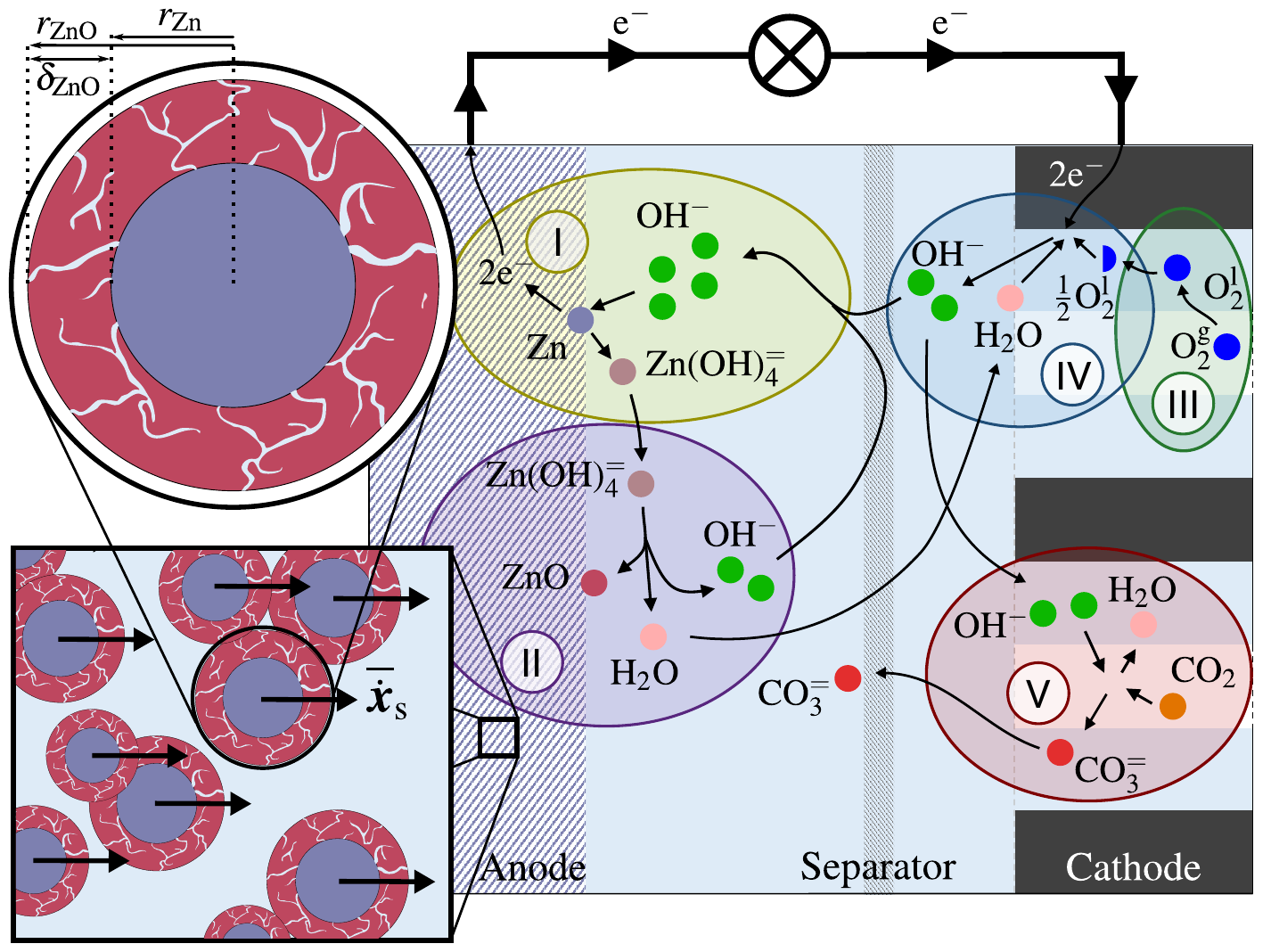}
	\caption[]{Reactions in the zinc-air cell: (I) zinc dissolution, (II) zinc oxide precipitation, (III) oxygen absorption into the electrolyte, (IV) oxygen reduction, (V) dissolution of atmospheric carbon dioxide and carbonate formation \cite{Stamm2017}. The zooms sketch the sub-model of the zinc anode. Spherical primitive particles are surrounded by a zinc core and a porous zinc oxide shell \cite{Horn2003}. The solid convection velocity $\lvaIn{\vel}_{\sol}$ moves these zinc particles. Figure similar to Ref. \cite{Stamm2017}}
	\label{fig:mod:lva:liq:schematics}
\end{figure}

In metal-air batteries, solid phases grow and dissolve during cycling. On the mesoscopic particle-scale, phase-field-like models can capture the dynamics of phase boundaries \cite{horstmann2013rate,Cogswell2015,Single2016,Single2017,Horstmann2019}. On the macroscopic cell-scale, more computationally efficient approaches would be necessary. We note that modern cell models follow the dynamics of volume fractions locally and neglect the non-local movement of moving phases \cite{Neidhardt2012}. In this case, the breathing of solid volume is just compensated by a local porosity change. However, operando tomographies like \cref{fig:es:tomography} show an additional convection in the $\zn$ anode because of different molar volumes $\molVol_{\zno}=1.58~\molVol_{\zn}$ \cite{Arlt2014}. We are the first to capture this phenomenon and model convection of solids in a cell model. 

We distinguish three phases that make up the total solid volume, i.e., $\epsS = \varepsilon_{\zn} + \varepsilon_{\zno} + \varepsilon_{\passive}$. The volume-averaged continuity equations for the solid phases are (see \cref{eq:mod:lva:con}) 
\begin{align}
\partial_{t} \lvaSup{1_{\zn}} = \partial_{t} \varepsilon_{\zn} &= -\bn \cdot  \left( \varepsilon_{\zn}\lvaIn{\vel}_{\sol} \right)  + \molVol_{\zn}  \lvaSurfC{\source}^{\Theta}_{\zn}  \, , \label{eq:mod:sa:eq_zn}\\
\partial_{t} \lvaSup{1_{\zno}} =\partial_{t} \varepsilon_{\zno} &= -\bn\cdot   \left( \varepsilon_{\zno}\lvaIn{\vel}_{\sol} \right) + \molVol_{\zno}  \lvaSurfC{\source}^{\Theta}_{\zno}  \, , \label{eq:mod:sa:eq_zno}\\
\partial_{t} \lvaSup{1_{\passive}} = \partial_{t} \varepsilon_{\passive} &= 0  \,,  \label{eq:mod:sa:eq_passive}
\end{align}
where the solid convection velocity $\lvaIn{\vel}_{\sol}$ applies to both phases consistent with our primitive particle model (see \cref{fig:mod:lva:liq:schematics}). In analogy to laminar Darcy flow of liquids through porous structures \cite{Horstmann2013}, it is calculated from a pseudo-pressure $\lvaIn{p}_{\sol}$  
\begin{align}
 \lvaIn{\vel}_{\sol} = \omega \bn \lvaIn{p}_{\sol}  \, , \label{eq:mod:sa:supSolVel}
\end{align}
with the pseudo-permeability $\omega$. 

Dependent on the solid saturation $s_{\sol} = {\left( \varepsilon_{\zn} + \varepsilon_{\zno} \right)}/{(1-\varepsilon_{\passive})}$, we determine the solid pseudo-pressure $\lvaIn{p}_{\sol}$ with a phenomenological Leverett function $J_{\sol}(s_{\sol})$, 
\begin{align}
	J_{\sol}(s_{\sol})  = \lvaIn{p}_{\sol}-\lvaIn{p}_{\gas}  \, . \label{eq:mod:sa:pressSolid}
\end{align}
We choose a Leverett function $J_{\sol}(s_{\sol})$ with a step at the critical saturation $s^{*}_{\sol}$. If the saturation is small, $s_{\sol}<s^{*}_{\sol}$, the Leverett function vanishes, $J_{\sol}(s_{\sol})=0$. In this first case, the solids grow locally and the solid convection vanishes, i.e., $ \lvaIn{\vel}_{\sol}=0$. However, if the saturation exceeds the critical saturation, $s_{\sol}>s^{*}_{\sol}$, the Leverett function increases with the saturation $s_{\sol}$. In this second case, the pseudo-pressure pushes the solids away and the solid convection is non-zero, i.e., $\lvaIn{\vel}_{\sol}\ne 0$. A mathematical description of $J_{\sol}(s_{\sol})$ is given in the Supplementary Materials in \cref*{sec:suppl:params:levs}. Note that alternative approaches to model $\lvaIn{p}_{\sol}$ are available \cite{Single2016,Single2017}. 

Now, we give details on our microscopic zinc particle model. X-ray/SEM measurements show that a zinc oxide shell forms around the zinc particles \cite{Horn2003}. We assume spherical $\zn$ particles surrounded by a porous $\zno$ shell as shown in \cref{fig:mod:lva:liq:schematics}. Our model solves for the density of primitive particles $\ppd(x,t)$, too. As primitive particles cannot emerge or vanish, they move with the solid convection velocity only, 
\begin{align}
\partial_{t} \ppd = -\bn \cdot \left( \ppd  \lvaIn{\vel}_{\sol} \right)  \, . \label{eq:mod:sa:eq_ppd}
\end{align}
According to \cref{fig:mod:lva:liq:schematics}, the $\zn$ radius $r_{\zn}$ and the $\zno$ radius $r_{\zno}$ are then calculated from 
\begin{align}
\varepsilon_{\zn}&=\ppd\cdot 4\pi r_{\zn}^3/3 ~,\\
\varepsilon_{\zno}&=\ppd\cdot \epsF\cdot 4\pi \left(r_{\zno}^3-r_{\zn}^3\right)/3~.
\end{align} 
Here, $\epsF=0.95$ is the solid volume fraction of the $\zno$ shell. Further details are given in the Supplementary Materials (see \cref*{sec:suppl:params:sa}).

Electrons are conducted by the solids in anode and cathode. The spatially-resolved electrode potential is determined by
\begin{align}
	0 = \bn \cdot \lvaSup{\myvec{j}_{\sol}} + F\lvaSurfC{\source}^{\Theta, \sol}_{\mathrm{q}} \, , \label{eq:mod:lva:sol:lvaSup_transport_eq_j_sol_ins}
\end{align}
with the volume-averaged flux term
\begin{align}
\lvaSup{\myvec{j}_{\sol}} \approx - \lvaIn{\kappa}_{\sol} \epsS \mytens{T}^{\mathrm{eff}}_{\sol} \cdot \bn  \cdot \lvaIn{\phi}_{\sol}\, . \label{eq:mod:lva:sol:mean_transport_terms}
\end{align}
The effective transport tensor $\mytens{T}^{\mathrm{eff}}_{\alpha}$ has to be modeled (see for example Refs. \cite{Whitaker1999,Weissberg1963,Quintard1993}). We model a macroscopic and isotropic medium which reduces $\mytens{T}^{\mathrm{eff}}_{\alpha}$ to a scalar tortuosity factor 
\begin{align}
	\mytens{T}^{\mathrm{eff}}_{\alpha} &\approx \frac{1}{\tau_{\alpha}} \mytens{I} = \varepsilon^{\bruggemann-1}_{\alpha} \mytens{I}  \, ,
	\label{eq:mod:lva:liq:isotropic_eff_transport_tensor}
\end{align}
with the Bruggemann coefficient $\bruggemann$ and the identity matrix $\mytens{I}$. 

The parameters for solid phase transport are stated in the Supplementary Materials in \cref*{sec:suppl:params:solid} \cite{Stamm2017,Horn2003,Manke2007,Miller1941}. 

\subsection{Liquid Phase Transport}
\label{sec:mod:lva:liq}
In this subsection, we summarize the consistent multi-species electrolyte model developed by Stamm et al. \cite{Stamm2017}. Our transport theory ensures a strictly positive entropy production and is thermodynamic consistent \cite{Latz2011,Latz2015}. Since we describe transport in the center-of-mass frame and consider the electro-neutrality of the electrolyte on macroscopic length scales, it suffices to calculate the motion of anions ($\oh$, $\znoh$, $\co$) and dissolved oxygen $\oli$. We state \cref{eq:mod:lva:con} for the independent concentrations $c_{\gl}$ and the vanishing charge density $\densChargeL=0$ in the porous electrode and separator, 
\begin{align}
\partial_{t} \lvaSup{c_{\glm}} = \partial_{t} \left(\epsL \lvaIn{c}_{\glm}\right) &= -\bn\cdot \lvaSup{\ncflux_{\glm}} - \bn\cdot \lvaSup{c_{\glm}\velL} + \lvaSurfC{\source}^{\Theta}_{\glm}   , \label{eq:mod:lva:lvaSup_transport_eq_c_ins_neg}\\
\partial_{t} \lvaSup{c_{\gln}} = \partial_{t} \left(\epsL \lvaIn{c}_{\gln}\right) &= -\bn\cdot \lvaSup{\ncflux_{\gln}} - \bn\cdot \lvaSup{c_{\gln}\velL} + \lvaSurfC{\source}^{\Theta}_{\gln}  \, , \label{eq:mod:lva:lvaSup_transport_eq_c_ins_neu}\\
0 = \partial_{t} \lvaSup{\densChargeL} &= -\bn\cdot \lvaSup{\myvec{j}_{\liq}}  + F\lvaSurfC{\source}^{\Theta, \liq}_{\mathrm{q}}   \, , \label{eq:mod:lva:liq:lvaSup_transport_eq_j_liq_ins}
\end{align}
The volume-averaged center-of-mass flux densities for ions $\ncflux_{\glm}$, neutral solutes $\ncflux_{\gln}$, and charges $\myvec{j}_{\liq}$ are
\begin{align}
\label{eq:mod:lva:liq:mean_transport_terms}
\lvaSup{\ncflux_{\glm}} & \approx - \lvaIn{D}_{\glm} \epsL \mytens{T}^{\mathrm{eff}}_{\liq} \cdot \bn \lvaIn{c}_{\glm}  + \frac{\lvaIn{t}_{\glm}}{z_{\glm} F} \lvaSup{\myvec{j}_{\liq}} \, , \\
\lvaSup{\ncflux_{\gln}} & \approx - \lvaIn{D}_{\gln} \epsL \mytens{T}^{\mathrm{eff}}_{\liq} \cdot \bn \lvaIn{c}_{\gln}  \, , \\
\lvaSup{\myvec{j}_{\liq}} &\approx - \lvaIn{\kappa}_{\liq} \epsL \mytens{T}^{\mathrm{eff}}_{\liq} \cdot \bn \lvaIn{\phi}_{\liq} - \frac{\lvaIn{\kappa}_{\liq}}{F} \epsL \mytens{T}^{\mathrm{eff}}_{\liq} \cdot \sum\limits_{\gl} \frac{\lvaIn{t}_{\gl}}{z_{\gl}}  \frac{\partial \lvaIn{\mu}_{\gl}}{\partial \lvaIn{c}_{\gl}} \bn \lvaIn{c}_{\gl} \, .
\end{align}
We use the electric potential $\phi_{\liq}$, diffusion coefficients $D_{\gl}$, electrolyte conductivity $\kappa_{\liq}$, the transference numbers $t_{\glm}$, chemical potentials $\mu_{\gl}$, and the charge numbers $z_{\gl}$. We approximate the averaged coefficients as $\lvaIn{D(c)} \approx D(\lvaIn{c})$. The effective transport tensor for an isotropic medium is given in \cref{eq:mod:lva:liq:isotropic_eff_transport_tensor}. The flux densities of $\water$ and $\kalium$ in the center-of-mass frame follow from the frame condition and the electrical current
\begin{align}
\sum\limits_{\bl} \molMass_{\bl} \lvaSup{\ncflux_{\bl}} \equiv 0, \hspace{0.5cm} 
\sum\limits_{\bl} z_{\bl} F \lvaSup{\ncflux_{\bl}} = \lvaSup{\myvec{j}_{\liq} }
\end{align}
with the partial molar volumes $\molMass_{\bl}$.

We take into account the volumetric electrolyte equation of state \cite{Horstmann2013}
\begin{align}
\lvaSup{1_\liq} = \sum\limits_{\bl} \lvaSup{\pmv_{\bl}c_{\bl}} \approx \sum\limits_{\bl} \lvaIn{\pmv}_{\bl} \lvaSup{c_{\bl}}  \, \label{eq:mod:bulk:incompr_constr}
\end{align}
with the partial molar volumes $\lvaIn{\pmv_{\bl}(c)}\approx \lvaIn{\pmv}_{\bl}(\lvaIn{c})$. For pressure-independent partial molar volumes, \cref{eq:mod:bulk:incompr_constr} realizes a multi-component incompressibility constraint. As shown in Ref. \cite{Stamm2017}, \cref{eq:mod:bulk:incompr_constr} yields the temporal evolution of porosity $\epsL$ in electrodes and separator,
\begin{align}
\partial_t\lvaSup{1_\liq}=\partial_{t} \epsL = -\bn\cdot\lvaSup{\velL} - \sum\limits_{\bl} \lvaIn{\pmv}_{\bl} \bn\cdot \lvaSup{\ncflux_{\bl}}   + \sum\limits_{\bl} \lvaIn{\pmv}_{\bl} \lvaSurfC{\source}_{\bl}^{\Theta} \, . \label{eq:mod:lva:liq:lvaSup_gae}
\end{align}
We note that \cref{eq:mod:lva:liq:lvaSup_gae} generalizes the standard incompressibility condition ${\bn\cdot \velL = 0}$ known for single-component and single-phase systems.  \cite{Latz2015}.

For realistic discharge rates of zinc-air batteries, the flow of the alkaline electrolyte is laminar. Hence, the center-of-mass velocity ${\lvaIn{\vel}_{\liq}}$ is the gradient of a scalar potential. We apply Darcy's law and determine the convective flux density $\lvaSup{c_{\gl}\velL}$ with the liquid pressure $\lvaIn{p}_{\liq}$ \cite{Horstmann2013}, 
\begin{align}
\lvaSup{c_{\gl}\velL} \approx \lvaIn{c}_{\gl}   \epsL \lvaIn{\vel}_{\liq} = - \epsL\lvaIn{c}_{\gl}  \frac{s_{\liq}B_{\liq}^{\mathrm{in}}}{\lvaIn{\eta}_{\liq}} \bn \lvaIn{p}_{\liq}  \, , \label{eq:mod:lva:liq:darcy_law}
\end{align}
where $B_{\liq}^{\mathrm{in}}$ is the intrinsic permeability and $\lvaIn{\eta}_{\liq}$ is the intrinsic viscosity of the liquid phase. The liquid saturation $s_{\liq} = {\epsL}/{(1 - \epsS)}$ serves as relative permeability and takes into account the multi-phase character of the system \cite{Muskat1936}. The liquid pressure $\lvaIn{p}_{\liq}$ is determined by the phenomenological Leverett function $J_{\liq}(s_{\liq})$ \cite{Danner2016,Danner2014}. It holds 
\begin{align}
J_{\liq}(s_{\liq}) = \lvaIn{p}_{\mathrm{c}} = \lvaIn{p}_{\gas} - \lvaIn{p}_{\liq}  \, ,
\label{eq:mod:lva:liq:levertt_function}
\end{align}
with the capillary pressure $\lvaIn{p}_{\mathrm{c}}$. 

The corresponding transport parameters are given in the Supplementary Materials in \cref*{sec:suppl:params:electrolyte} \cite{Siu1997,Stamm2017,Zen1957,Bahita1968,May1978,Cents2003,Zeebe2011,Davis1967,See1997,Liu1981_2,Dean1999,Sunu1980,Newman2004,Zen1957,Bear1972,Siu1997,Danner2014,Danner2016}.

\subsection{Gas Phase Transport}
\label{sec:mod:lva:gas}
In a zinc-air button cell, the gas phase is present in anode, separator, and cathode. The gaseous volume fraction $\epsG$ follows from the constraint on phase volumes in \cref{eq:mod:lva:sum_volume_fraction} as $\epsL$, $\varepsilon_{\zn}$, $\varepsilon_{\zno}$, and $\varepsilon_{\passive}$ are known from \cref{eq:mod:lva:liq:lvaSup_gae}, \cref{eq:mod:sa:eq_zn}, \cref{eq:mod:sa:eq_zno}, and \cref{eq:mod:sa:eq_passive}, respectively. 

As shown in Ref. \cite{Horstmann2013}, the partial pressures in the gas phase remain almost constant during discharge of metal-air batteries with aqueous alkaline electrolytes. Therefore, we keep the partial gas pressures $\lvaIn{p}_{\og}$ and $\lvaIn{p}_{\cotwo}$ constant. The total gas pressure is the atmospheric pressure $\lvaIn{p}_{\gas}=p_{\mathrm{atm}}^{\ast}$.

\subsection{Reaction Rates}
\label{sec:mod:reac}
We depict the reactions in the zinc-air cell in \cref{fig:mod:lva:liq:schematics} and denote them with the indices $\mathcal{I} = \left\lbrace \mathrm{I}, \mathrm{II}, \mathrm{III}, \mathrm{IV}, \mathrm{V} \right\rbrace$. The total surface source terms $\lvaSurfC{\source}_{\beta_{\alpha}}^{\Theta}$ are calculated as sums of products of the total reaction rates per surface area $j_{i}$ with their specific surface areas ${\abs{A_{\alpha, \alpha'}}}/{\abs{V}}$, 
\begin{align}
	\lvaSurfC{\source}_{\beta_{\alpha}}^{\Theta} =  \sum\limits_{\alpha'} \frac{\abs{A_{\alpha, \alpha'}} }{|V|}\sum\limits_{i \in \mathcal{I}} \chi^{\beta_{\alpha}}_i j^{\alpha, \alpha'}_{i}  \,  \label{eq::mod:lva:reac:surf_source_term},
\end{align}
with the stoichiometric coefficients $\chi^{\beta_{\alpha}}_i$ of species $\beta_{\alpha}$ in reaction $i$. The sources for ionic charge $\lvaSurfC{\source}_{\mathrm{q}}^{\Theta,\liq}$ and $\lvaSurfC{\source}_{\mathrm{q}}^{\Theta,\sol}$ are
\begin{align}
\lvaSurfC{\source}_{\mathrm{q}}^{\Theta,\liq}=-	\lvaSurfC{\source}_{\mathrm{q}}^{\Theta,\sol} = \sum\limits_{\alpha'} \frac{\abs{A_{\alpha, \alpha'}}}{|V|}  \sum\limits_{i \in \mathcal{I}} \chi^{e^-}_i  j_{i}  \, ,
	\label{eq::mod:lva:reac:surf_source_term_q}
\end{align}
with the stoichiometric coefficients $\chi^{e^-}_i$ of electrons $e^-$ in reaction $i$. 

The individual reaction rates $j_{i}$ are stated in the Supplementary Materials (see \cref*{sec:suppl:params:reac}). Note that we summarize carbon dioxide absorption and carbonate formation in a single surface reaction as shown by Stamm et al. \cite{Stamm2017}. Thus, all volume source terms vanish in our model. 

We introduced our microscopic model of spherical particles in \cref{sec:mod:lva:sol}. The $\zno$ shell around the $\zn$ core hinders the ion transport from the bulk electrolyte to the $\zn$ surface. Hydroxide is the main reactant in zinc dissolution at the $\zn$ surface. Thus, we model the diffusion of hydroxide through the $\zno$ shell assuming spherically symmetric $\zn$ particles and $\zno$ shells (see \cref{fig:mod:lva:liq:schematics}). We calculate the surface concentration $c_{\mathrm{s}, \oh}$ by a radial diffusion from the bulk to the $\zn$ surface \cite{Stamm2017}
\begin{align}
\left( 4- 2\epsF \frac{\molVol_{\zn}}{\molVol_{\zno}} \right) j_{\mathrm{I}} = \left( 1 - \epsF \right)^{3.5} \lvaIn{D}_{\oh} \frac{ \lvaIn{c}_{\oh} - c_{\mathrm{s}, \oh} }{\delta_{\zno}} \frac{r_{\zno}}{r_{\zn}} \, , \label{eq:mod:lva:liq:surface_conc_oh}
\end{align}
with the reaction rate $j_{\mathrm{I}}$, its thickness $\delta_{\zno}=r_{\zno}-r_{\zn}$, and the inner and outer radius of the passivation layer $r_{\zn}$ and $r_{\zno}$.  The elevated Bruggemann coefficient is motivated by low porosities of the $\zno$ shell $1-\epsF=0.05$. The factor $\left( 4- 2\epsF \molVol_{\zn}/\molVol_{\zno} \right)$ is the number of hydroxide ions passing the $\zno$ shell per dissolved zincate ion. Here, we assume that the dissolved $\zn$ phase is completely replaced by $\zno$. Thus, hydroxide ions are created below the passivation shell, too.

Details on reactions and transport parameters are stated in the Supplementary Materials in \cref*{sec:suppl:params:reac} \cite{Latz2013,Bazant2013,Stamm2017,Sunu1980,Dean1999,Deiss2002,Sunu1978,Horstmann2013,Liu1981_1,Liu1981_2,Eames1997,Weisenberger1996,Sander2015,Eberle2014,Gyenge2012,Pohorecki1988,TIBTECH}.

\section{Computational Details}
\label{sec:cd}

\subsection{Parameters}
\label{sec:cd:para}
Parameters represent the Renata ZA13 button cell \cite{RENATA_1}. Solid phase, liquid phase, and reaction parameters for the aqueous alkaline electrolyte (32 weight percent KOH) are discussed by Stamm et al. \cite{Stamm2017}. They are stated in the Supplementary Materials in \cref*{sec:suppl:params:solid}, in \cref*{sec:suppl:params:electrolyte}, and in \cref*{sec:suppl:params:reac}, respectively.  

Regarding the structure of the zinc electrode during cycling, the initial radius of the primitive zinc particles was measured with x-ray tomography \cite{Franke-Lang2017}, the formation of zinc oxide shells was observed with scanning electron microscopy \cite{Horn2003}, and the solid phase movement during discharge was imaged with x-ray tomography \cite{Arlt2014,Schroder2014_2}. We model the dissolution and growth of primitive {$\zn$}/{$\zno$} particles accordingly (see \cref*{sec:suppl:params:sa}). Our phenomenological Leverett function $J_{\sol}(s_{\sol})$ is adjusted to yield the observed solid convection during discharge (see \cref*{sec:suppl:params:levs}). 

\subsection{Numerics}
\label{sec:cd:num}
Our in-house tool, Battery and Electrochemistry Simulation Tool (BEST), provides a framework for multi-dimensional finite-volume discretized electrochemical simulations \cite{Latz2011,Latz2015}. BEST was originally developed at Fraunhofer ITWM for simulation of micro-structures on the meso-scale. We employ a backward differentiation formula for the time derivative and a Crank-Nicolson step for the time integration \cite{Quarteroni2012}. Convection is implemented with an upwind scheme \cite{Quarteroni2012}. The resulting nonlinear systems of equations are solved with an iterative Newton method.

We perform 2D simulations on a radially symmetric cylindrical grid which resembles the button cell geometry. We use a resolution of $(70, 3, 9) \times 20$ ((anode, separator, cathode) x-dir $\times$ radial discretization units) in \cref{sec:rse:val} and $(50, 1, 3) \times 12$ in all other sections. Our resolutions are sufficiently accurate as demonstrated in the Supplementary Materials in \cref*{sec:suppl:ad}. In \cref*{sec:suppl:val1D}, we compare our new 2D simulations with the existing 1D code used by Stamm et al. \cite{Stamm2017} and demonstrate accurate implementation.

The multi-component incompressibility constraint in \cref{eq:mod:lva:liq:lvaSup_gae} stiffens the system of equations. We improve the numerical time integration to yield stable simulations. A detailed description and error analysis of this algorithm will be presented in an upcoming publication. 

\subsection{Construction of Initial Data}
\label{sec:cd:init}
After measurements, tomographic images are reconstructed using the Octopus software \cite{Vlassenbroeck2007}. Firstly, we subtract the passive cell housing. Secondly, we apply a median filter to the reconstructed data (8 Bit gray values) reducing measurement noise. Thirdly, we perform segmentation of different phases inside the 3D volume by gray value thresholding. Finally, we add the passive cell housing again. The reconstructed data has a resolution of $491 \times 816 \times 816$ pixels and is segmented, i.e., each voxel is associated to one phase. 

Gray value thresholding relies on different beam attenuation of different phases. Dark pixels have less beam attenuation, which correspond to gas volumes. Bright values mean high absorption, which represents {$\zn$} here. Electrolyte and {$\zno$} have similar medium attenuation coefficients, that is why they are hardly separable. A better segmentation could be achieved by stochastic methods using Fiji tools and WEKA segmentation \cite{Schindelin2012,Arganda-Carreras2017}. The gray value thresholds are determined at all state-of-charges from the expected {$\zn$} and {$\zno$} amounts. By assuming that zinc is completely transformed to zinc oxide at the end-of-discharge, we estimate the initial amount of {$\zn$} as \SI{5.59e-3}{\mole} and of {$\zno$} as \SI{6.14e-4}{\mole}. From the discharge conditions in Ref. \cite{Arlt2014}, we then calculate the amounts of {$\zn$} and {$\zno$} at each state-of-charge. We perform simulations with and without {$\zno$} in order to determine if {$\zno$} is present in a fresh cell (see \cref{sec:rse:val}). The resulting threshold values are presented in \cref{tab:cd:init:thresholds}.

\begin{table}[!tb]
	\centering
	\makegapedcells
	\small
	\begin{tabular}{l|r r r}
		\toprule[2pt]
		Phase & \  \SI{100}{\percent} SOC & \  \SI{47}{\percent} SOC & \  \SI{0}{\percent} SOC \\ 
		\midrule[1.5pt]
		gas & 65,5 & 65,5 & 65,5 \\
		\hline
		$\zno \ (\varepsilon^{0}_{\zno} = 0)$ & 121,5 & 102,5 & 104,5 \\
		\hline
		$\zno \ (\varepsilon^{0}_{\zno} \neq 0)$ & 110,5 & 99,5 & 101,5 \\
		\hline 
		$\zn$ & 121,5 & 133,5 & 141,5 \\
		\hline 
		passive material & 200,5 & 200,5 & 200,5 \\ 
		\bottomrule[2pt]
	\end{tabular}
\normalsize
\caption[Threshold values for reconstructed data segmentation.]{Threshold values for reconstructed data (8 Bit) segmentation. To match the correct total mass of zinc and zinc oxide, we adapt the threshold values in each measurement.}
\label{tab:cd:init:thresholds}
\end{table}

We construct the 2D initial data for the simulations by volume-averaging the 3D reconstructed data. The volume-average is performed in cylindrical coordinates, including a radial average. We average all phases within the constructed discretization units (see \cref{fig:es:tomography,fig:rse:val:volFracs_compExpSim}).

\section{Validation and Results}
\label{sec:rse}
In this section, we present 2D simulation results for the discharge and charge of zinc-air button cells. Firstly, we validate our 2D simulations by comparing with measurements during cell discharge (see \cref{sec:rse:val}). Secondly, we evaluate the effect of electrolyte carbonation on cycle-life (see \cref{sec:rse:cycCarb}). Finally, we simulate cycling at various current densities (see \cref{sec:rse:vcd}).

\subsection{Validation During Cell Discharge}
\label{sec:rse:val}
\begin{figure*}[t]
    \centering
    \includegraphics[width=0.95\textwidth]{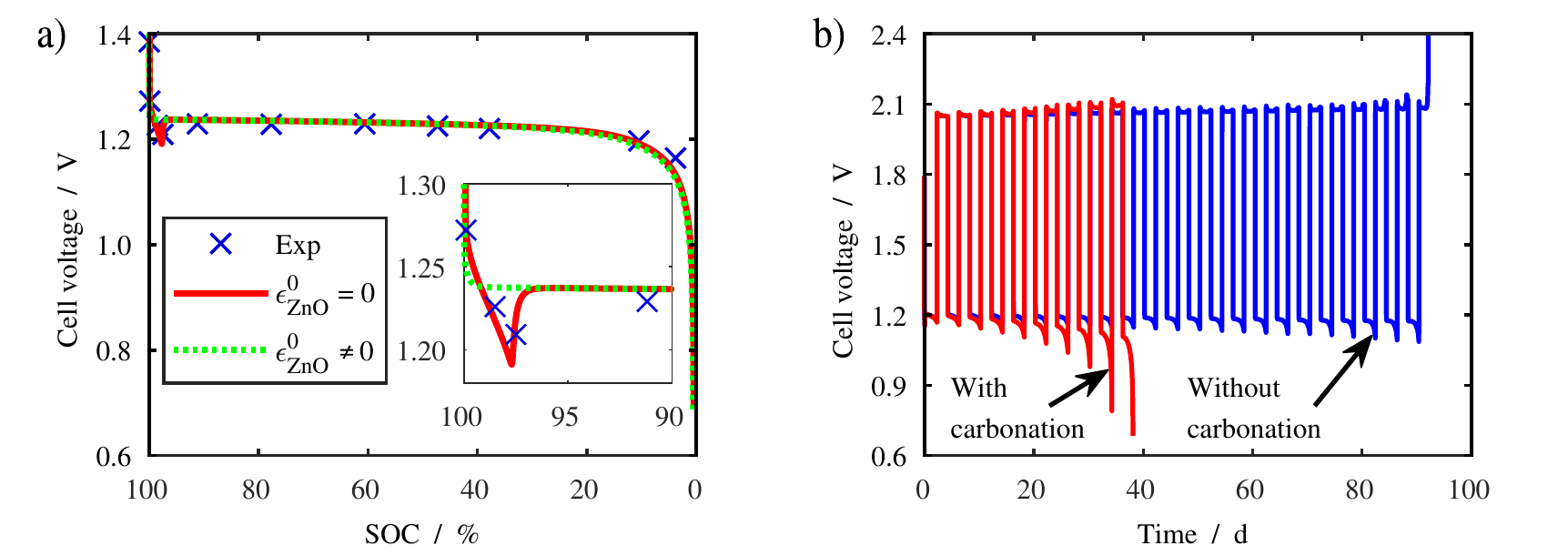}
    \caption[Validation of cell voltage and influence of carbonation.]{Cell voltage during discharge and cycling. (a) Validation of simulated discharge voltage with experiments (blue crosses). We initialize the two simulations with (dashed green line) and without (red line) zinc oxide (${T = \SI{293.15}{\kelvin}}$; ${\myvec{j} = \SI{54.2}{\ampere \per \meter\squared}}$). The inset shows that the experiments suggest the absence of initial $\zno$. (b) Effect of electrolyte carbonation on life-time during cycling. We compare the cell voltage with (red line) and without (blue line) $\cotwo$ in the feed gas. Electrolyte carbonation reduces electrolyte conductivity causing zincate depletion and end-of-life at $\sim\SI{92}{\day}$ (${T = \SI{298.15}{\kelvin}}$; ${\myvec{j} = \SI{54.2}{\ampere \per \meter\squared}}$). }
    \label{fig:rse:val:cellVoltage}
\end{figure*}
\begin{figure*}[!t]
	\centering
		\includegraphics[width=\textwidth]{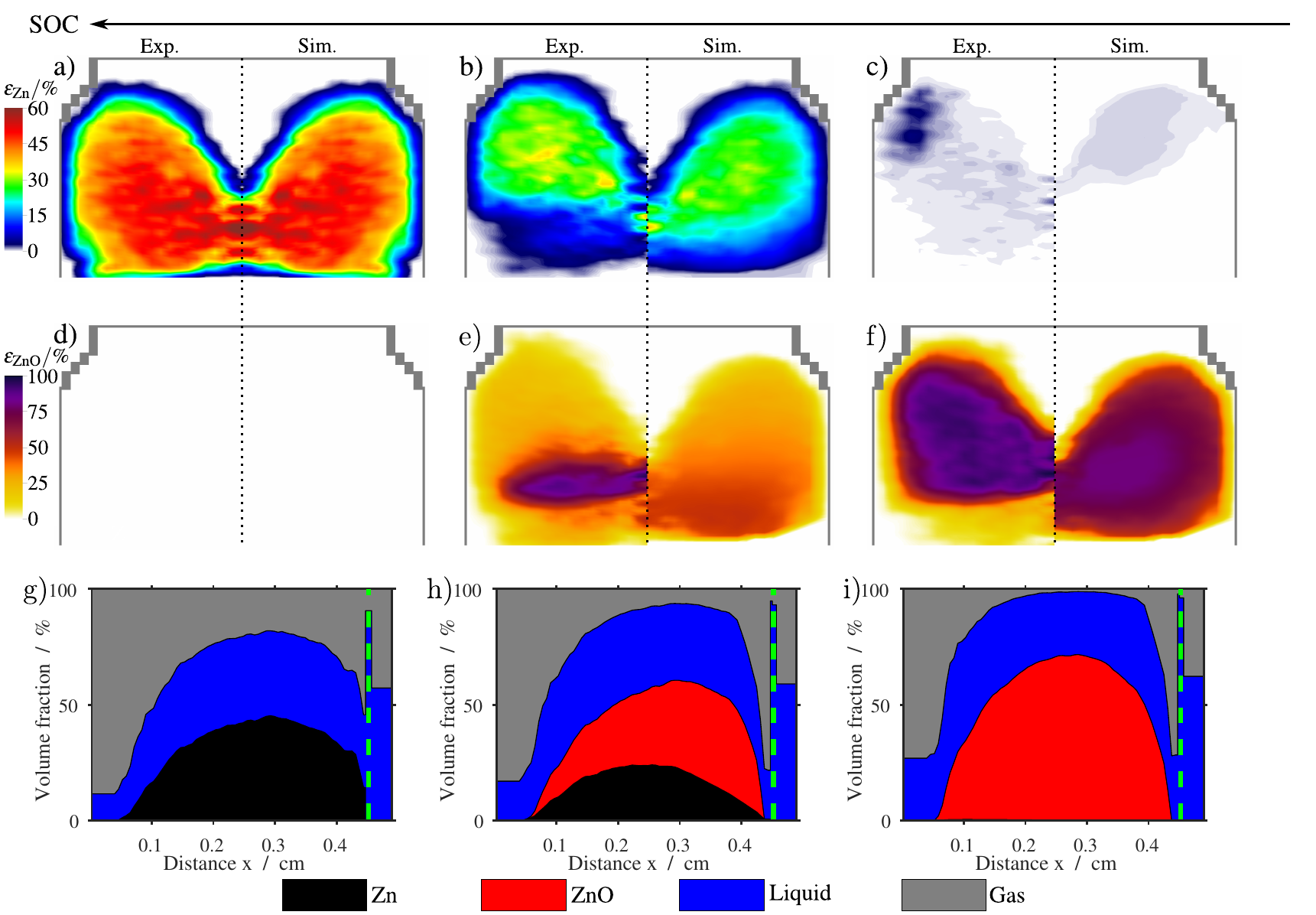}
	\caption[Validation of the solid volume fractions during discharge.]{Validation of the solid volume fractions during discharge for ${\varepsilon^{0}_{\zno} = 0}$. The state-of-charge is decreasing from \SI{100}{\percent} to \SI{47}{\percent} to \SI{0}{\percent} from left to right. The left halves of the cells in (a-f) show volume-averaged tomographies. The right halves are simulation results. Separator and cathode are located directly beneath the plotted anode. The cell housing is plotted in gray. Figures (g-i) are radial projections of the simulated phase distribution onto the through direction (x-direction). The dashed green line represents the separator position (${T = \SI{293.45}{\kelvin}}$; ${\myvec{j} = \SI{54.2}{\ampere \per \meter\squared}}$).}
	\label{fig:rse:val:volFracs_compExpSim}
\end{figure*}

For model validation during galvanostatic discharge, we compare our simulation results with experimentally measured cell voltages and tomographic images (see \cref{fig:rse:val:cellVoltage}a and \cref{fig:rse:val:volFracs_compExpSim}). The voltage curve contains information about $\zno$ nucleation and growth. X-ray tomography reveals the dynamics of $\zn$ electrode morphology.

It is debated whether $\zno$ is admixed to the anode initially. In Ref. \cite{Arlt2014}, simulations were initialized with a small amount of solid $\zno$; in Ref. \cite{Stamm2017}, we concluded that the analyzed primary zinc-air button cell contains a pure $\zn$ anode initially. Therefore, we carry out two simulations. Based on tomographic images of the fresh cell, we initialize one with $\varepsilon^{0}_{\zno} = 0$ and the other one with $\varepsilon^{0}_{\zno} \neq 0$ (see \cref{tab:cd:init:thresholds}). 

\cref{fig:rse:val:cellVoltage} compares measured and simulated discharge voltages. After $\text{SOC} = 90\%$, both simulations and the experiment agree nicely, apart from a minor deviation at the end-of-discharge. The voltage curve is relatively flat as expected for a conversion-type battery chemistry \cite{Stamm2017}. At the beginning of discharge, however, the two simulations deviate from each other. The simulation with admixed $\zno$ predicts a dip of $\sim \SI{50}{\milli\volt}$ which is absent in the simulations without admixed $\zno$. The inset of \cref{fig:rse:val:cellVoltage}a highlights that the experiment shows a dip in cell voltage. State-of-charge and voltage amplitude of this dip very well agree to the predictions of the simulation with $\varepsilon^{0}_{\zno} = 0$.

As discussed by Stamm et al. \cite{Stamm2017}, the voltage dip indicates $\zno$ nucleation. If $\zno$ is not admixed to the $\zn$ paste, i.e., $\varepsilon^{0}_{\zn} = 0$, a critical $\znoh$ supersaturation has to be reached before $\zno$ can continuously precipitate. The resulting peak in the $\znoh$ concentration is reflected in the dip in cell voltage. Thus, we choose a pure $\zn$ paste without admixed $\zno$ as initial condition for the simulations in the remainder of this section.  

For further validation, we compare the simulated $\zn$ and $\zno$ distribution within the anode with tomography data during galvanostatic discharge. In \cref{fig:rse:val:volFracs_compExpSim}, we show the phase distribution for the initial, the half-discharged, and the fully-discharged state. In \cref{fig:rse:val:volFracs_compExpSim}a-c, we observe in simulations and experiments that $\zn$ preferentially dissolves close to the separator. Therefore, at the end-of-discharge the remaining $\zn$ is located near the lid at the top of the button cell. $\zno$ forms inside the $\zn$ structure as shown in \cref{fig:rse:val:volFracs_compExpSim}d-f. The tomography records an inhomogeneous $\zno$ distribution at intermediate state-of-charges and an accumulation near the center of the cell. Our simulation results in a slightly flatter distribution of $\zno$ with a peak close to the separator. In the simulations, $\zno$ does not precipitate in a tiny region directly near the separator (see \cref{fig:rse:val:volFracs_compExpSim}g-i). The experiment, instead, reveals a larger region, with a nonzero, but significantly reduced $\zno$ volume fraction. The overall $\zno$ distributions of experiment and simulation have a qualitatively similar shape in the fully-discharged state. On closer inspection, experiments show a slight movement of $\zn$ and $\zno$ to the lid, which is absent in simulations. 

Let us begin with discussing the zinc distribution. We observe that our simulation is in very good agreement with the measurement. In both, simulation and experiment, we observe a slightly faster dissolution near the separator during discharge and some $\zn$ remaining near the lid at the end-of-discharge. This is caused by a hydroxide concentration gradient from the cathode to the anode. The slight movement to the top of the button cell in the experiment results from the larger molar volume of $\zno$ than of $\zn$. Since the anode is a $\zn$ paste without any supporting structure, this volume expansion induces a movement. The separator on the bottom acts as a boundary, while there is enough space near the lid. Thus, the overall movement of the solid phase is faced towards the lid, where the gas is compressed or extruded. Our novel model for solid convection in \cref{eq:mod:sa:supSolVel} does not reproduce this slight movement during a single discharge. It induces a solid phase movement only if the solid volume saturation rises above a certain threshold. During cycling, however, our model captures the continuous movement of solid phase as shown below in \cref{sec:rse:vcd}. We highlight that this is the first cell model capturing this effect.

With respect to the zinc oxide phase, simulations also predict the overall shape of the $\zno$ block correctly. Due to its negative charge, zincate drifts towards the lid. Therefore, $\zno$ shows a clear peak structure in experiment and simulation as highlighted in \cref{fig:rse:val:volFracs_compExpSim}h-i. Nevertheless, there are deviations to the tomographies in a detailed view. As our previous 1D simulations \cite{Stamm2017}, our 2D simulations predict a small region close to the separator, where no $\zno$ precipitates. The gap in this work is smaller than in our previous work because of a different initial phase distribution and a different parametrization of the nucleation model. We note that the size of this region depends on the details of the nucleation model. Our tomographic images do not show a region without any $\zno$ precipitation, but a large region with low $\zno$ volume fraction close to the separator. We note that the reconstructed data of $\zno$ is insufficient for a concluding comparison because $\zno$ and electrolyte phases cannot be clearly distinguished. Higher resolution tomography with electron microscopy would allow to improve the nucleation model and to unambiguously determine the phase distribution (compare Ref. \cite{Horn2003}).

To sum up the comparisons of the cell voltage profile and the tomography data, we find that or model is in a good agreement with the measurements, but that modeling of $\zno$ nucleation is challenging. The fresh primary cell contains a pure $\zn$ anode without admixed $\zno$. 

\subsection{Carbonation of Electrolyte During Cycling}
\label{sec:rse:cycCarb}
We simulate the effect of $\cotwo$ in the feed gas during battery cycling. In this case, $\cotwo$ is absorbed by the electrolyte and reacts to carbonate $\co$ \cite{Stamm2017}. In \cref{fig:rse:val:cellVoltage}, we compare simulations for ambient air with $\cotwo$ and synthetic air without $\cotwo$. After a first discharge to $\SI{21}{\percent}$ state-of-charge, we galvanostatically cycle the cell between $\SI{90}{\percent}$ and $\SI{21}{\percent}$ state-of-charge. In the case of carbonation, we observe a clear drop of performance from cycle to cycle. The charging voltage increases slightly, while the discharge voltage decreases significantly. After 10 cycles ($\sim 38$ days), the cell voltage diverges at the end of discharge and the simulation terminates. In the case of pure oxygen supply, the voltage levels remain almost constant. After 24 cycles ($\sim 92$ days) at the end of the charging, the cell voltage diverges.

In Ref. \cite{Stamm2017}, we discussed in detail how carbonation is limiting the shelf-life of alkaline zinc-air batteries during storage. Here, we summarize this effect for battery cycling. Electrolyte carbonation due to reaction (V) decreases the overall $\oh$ concentration and reduces electrolyte conductivity. Because $\oh$ is required for $\zn$ dissolution in reaction (I), the anode overpotential increases. Finally at the end of the 10th cycle, the supply of $\oh$ to the $\zn$ metal surface is insufficient to sustain the cell current and the cell voltage diverges.  Without carbon dioxide in the feed gas, electrolyte stability is improved. The slow continuous decrease in efficiency due to a reshaping of the anode is discussed in the subsequent \cref{sec:rse:vcd}. 

In order to evaluate long-term battery operation, we simulate cells with synthetic air without $\cotwo$ in the gas feed in the remainder of this article.

\subsection{Simulation of Cycle-Life at Various Current Densities}
\label{sec:rse:vcd}

\begin{figure*}[!tb]
\centering
\includegraphics[width=\textwidth]{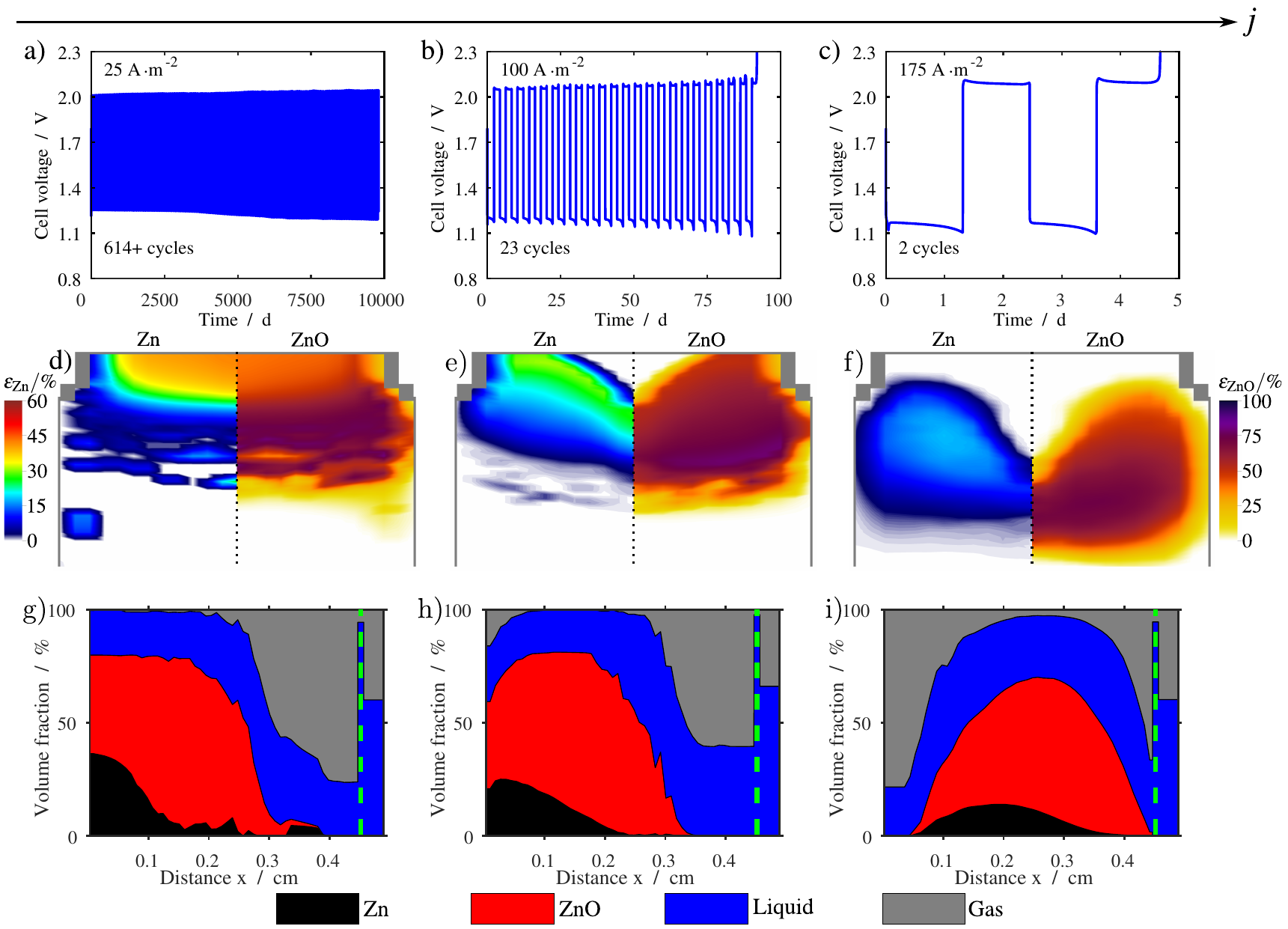}
\caption[Cell voltage and phase distribution under various currents.]{Cell voltage and phase distribution under various currents. Each column represents a different current density. (a)-(c) The first row shows the cell voltage over time. (d)-(f) The second row shows the phase distribution at the end of the simulation with $\zn$ on the left and $\zno$ on the right. (g)-(h) The phase distribution is plotted as a projection of the radial axis onto the through direction (connecting anode and cathode) in the third row. The dashed green line represents the separator. Cycling between $\SI{20}{\percent}$ and $\SI{90}{\percent}$ SOC (${T = \SI{298.15}{\kelvin}}$; ${\myvec{j} = \SI{54.2}{\ampere \per \meter\squared}}$).}
\label{fig:rse:vcd:currentVar}
\end{figure*}

Subsequent to model validation, we simulate repeated discharge and charge cycles of zinc-air cells for various current densities and analyze the zinc shape-change effect. Because ambient air with $\cotwo$ limits battery life-time \cite{Stamm2017}, we assume synthetic air without $\cotwo$ as feed gas. The initial $\zn$ and $\zno$ distribution in the cell is visualized in Figs. \ref{fig:rse:val:volFracs_compExpSim}a,d,g. It constitutes our reference for a quantification of the shape-change effect. The applied galvanostatic current densities range from $\SI{25}{\ampere \per \meter\squared}$ to $\SI{175}{\ampere \per \meter\squared}$. We cycle the cells between $\SI{21}{\percent}$ and $\SI{90}{\percent}$ state-of-charge. 

We show in \cref{fig:rse:vcd:currentVar}a-c that cycle-life is shorter if greater current densities are applied. Note that available computation time limits the simulation at the lowest current density $\SI{25}{\ampere \per \meter\squared}$ and we do not observe the full cycle-life. At the end-of-life, cell voltage diverges. During a single discharge, cell voltage generally decreases apart from the nucleation dip; during a single charge, cell voltage is U-shaped. The overpotential for charge and discharge is increasing from cycle to cycle.  

The cell voltage profile during a single discharging is explained above in \cref{sec:rse:val}. During charging, two concurrent effects result in the U-shaped voltage profile. On the one hand, deposition of $\zn$ increases $\zn$ volume fraction and specific surface area. This decreases the cell potential. On the other hand, dissolution of $\zno$ decreases $\zno$ volume fraction and specific surface area. Thus, $\znoh$ concentration in the electrolyte decreases. This increases the cell potential. During repeated discharge and charge cycles, solid phases move towards the lid and agglomerate as discussed below. Therefore, the increase in cell voltage towards the end-of-charge becomes more prominent from cycle to cycle. Cycle-life is ultimately limited by agglomeration of solid phases and depletion of $\znoh$ in the electrolyte.

The phase composition in the fully discharged state just before the end-of-life is shown in \cref{fig:rse:vcd:currentVar}d-i. We observe that $\zn$ and $\zno$ are shifted away from the cathode. For low current densities, the solids accumulate at the lid. For the greatest examined current density $\SI{175}{\ampere \per \meter\squared}$, they do not reach the lid. 

The loss of efficiency and limit of cycle-life is a consequence of the movement and reorganization of the solid phases. Firstly, the farther the solids move away from the cathode, the longer the transport paths in the electrolyte become. This increases the transport losses. The second, even more important, reason is the shrinkage of the total active surface area. Agglomeration of $\zno$ reduces the total surface area, which implies higher overpotentials. Eventually, the solids hit the housing. Finally, the total $\zno$ surface becomes too small to provide enough zincate ions during charging, as required by reaction I. Therefore, the zincate concentration depletes, leading to a diverging cell potential. For low enough current density, the zincate concentration does not deplete. The greater the current density, the earlier $\znoh$ depletion is reached (see \cref{fig:rse:vcd:currentVar}). For the large current density $\SI{175}{\ampere \per \meter\squared}$, the shrinking $\zno$ surface area causes a zincate depletion already after three cycles, even though the solid phases do not touch the lid.

\begin{figure}[!tb]
	\centering
	\includegraphics[width=0.95\columnwidth]{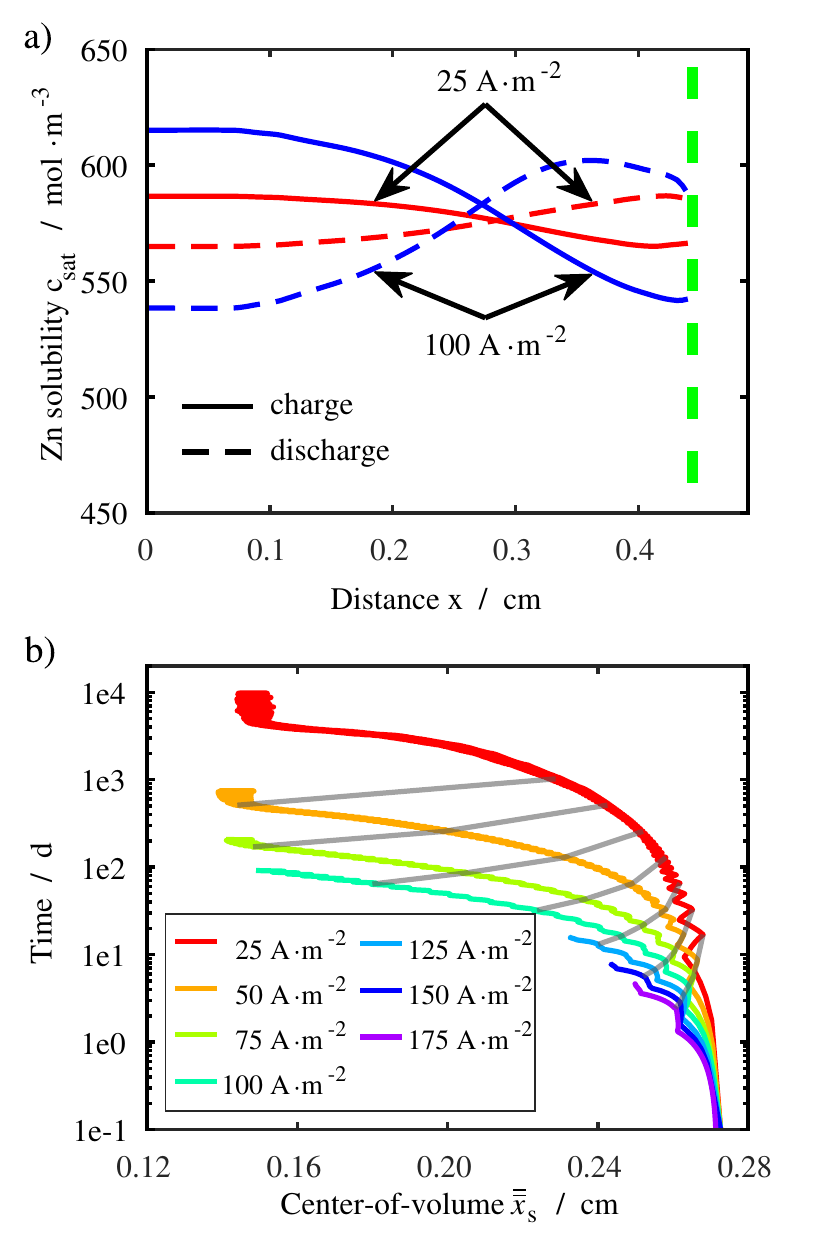}
	\vspace*{-3mm}
	\caption[Radial projection of the saturation concentration.]{(a) Examination of the solid phase movement by a radial projection of the $\znoh$ saturation concentration onto the x-direction. Higher current densities cause steeper gradients. The dashed green line represents the separator. (b) Movement of solids as center-of-volume movement during repeated discharge and charge. The gray lines represent times of constant cycle numbers. The marked cycle numbers have logarithmic spacing increasing from bottom right to top left (${T = \SI{298.15}{\kelvin}}$).}
	\label{fig:rse:vcd:currentVarConcSat}
\end{figure}

The observed zinc shape-change results from the relation of $\znoh$ concentration and solubility limit $\satConc$ as shown in \cref{fig:rse:vcd:currentVarConcSat}a. The $\znoh$ saturation concentration is inhomogeneous in through direction. Its profile is steeper for larger current densities. The sign of the gradient is different during charging and discharging. The saturation concentration is determined by the the potassium concentration in the electrolyte (see \cref*{eq:suppl:params:reac:II:c_sat}). During discharging, the solubility is lower at the lid than at the cathode and $\znoh$ accumulates at the lid due to the electric field. This leads to faster precipitation of $\zno$ near the lid (see reaction II). During charging, the saturation concentration is larger at the lid than at the cathode. A large solubility results in faster dissolution, providing more $\znoh$ near the lid. Thus, the $\zn$ deposition rate at the lid increases (see reaction I). The combination of both mechanisms tends to shift the solid phases away from the cathode, resulting in an overall $\zn$ and $\zno$ movement during cycling.

\cref*{fig:suppl:misc:movement} in the Supplementary Materials depicts the reaction rate of reaction (I). Positive values correspond to dissolution and negative values correspond to formation of $\zn$. We observe that $\zn$ dissolves and forms predominantly near the cathode. Comparing dissolution and deposition, however, $\zn$ formation occurs slightly closer to the lid than $\zn$ dissolution. Therefore, charging and discharging do not exactly balance each other and a phase-shift to the lid remains.

In order to examine the movement characteristics, we plot the center-of-volume for the solid phase in \cref{fig:rse:vcd:currentVarConcSat}b as defined in \cref*{eq:position}. It shows the central position of the active solid phases over time for different galvanostatic current densities. All simulations result in a zigzag course of the central position with a tendency towards the lid. Independent from the current, the center-of-volume remains above $\sim\SI{1.3}{\milli\meter}$. Simulations with high current densities show a termination already before the $\zno$ reaches the cell housing. Higher current densities cause a faster shift of the solid phases in the same number of cycles.

An important observation is the zigzag course of the center-of-solid-volume. This implies that during charging and discharging the shift has partially opposite directions. However, it does not cancel each other, but the discharge outweighs with a movement towards the lid. We also find that the backwards shift during charging is more dominant for lower currents. Using this characteristic helps to reduce the shape-change with an adjusted charge-discharge protocol. A detailed analysis on this is given in \cref{sec:ic:cdp}.

The analysis of the central position of the active solid phases in \cref{fig:rse:vcd:currentVarConcSat}b agrees with the results in \cref{fig:rse:vcd:currentVar}. A higher current causes faster shape-change and reduced cycle-life. At some point, the movement is limited because the block of solids hits the lid. As shown in \cref{fig:rse:vcd:currentVar}, the solids do not reach the cell housing for high current densities. In this case, agglomeration occurs in the center of the cell as shown in the Supplementary Materials in \cref*{fig:suppl:currentVarConcSat}.

We summarize that $\zn$ electrode shape-change limits cycle-life of zinc-air batteries and that lower cell currents significantly reduce shape-change.

\section{Discussion of Cycle-Life}
\label{sec:do}
We discuss strategies to reduce the zinc shape-change effect and improve cycle-life. In \cref{sec:ic:do}, we evaluate various $\zn$ anode morphologies including 3D current collectors and inhomogeneous $\zn$ distributions. The cycling protocol has a very strong influence on cycle-life as discussed in \cref{sec:ic:cdp}. Finally, in \cref{sec:ic:ft}, we analyze how better reaction and transport kinetics could improve cycle-life.

\subsection{Zinc Electrode Design Variation}
\label{sec:ic:do}
Here, we discuss how variations of the initial distribution of the solids in the anode affect cycle-life. We use the Renata ZA13 cell \cite{RENATA_1} as reference. Its initial phase distribution is shown in Figs. \ref{fig:rse:val:volFracs_compExpSim}a,d,g. We simulate 23 cycles for a galvanostatic cycling current density of $\SI{100}{\ampere \per \meter\squared}$ (see \cref{fig:rse:vcd:currentVar}b). We add $2\text{ volume percent}$ $\zno$ to the $\zn$ paste for avoiding inhomogeneous $\zno$ nucleation as demonstrated in \cref{sec:rse:val}.

\begin{figure*}[t!]
    \centering
    \includegraphics[width=\textwidth]{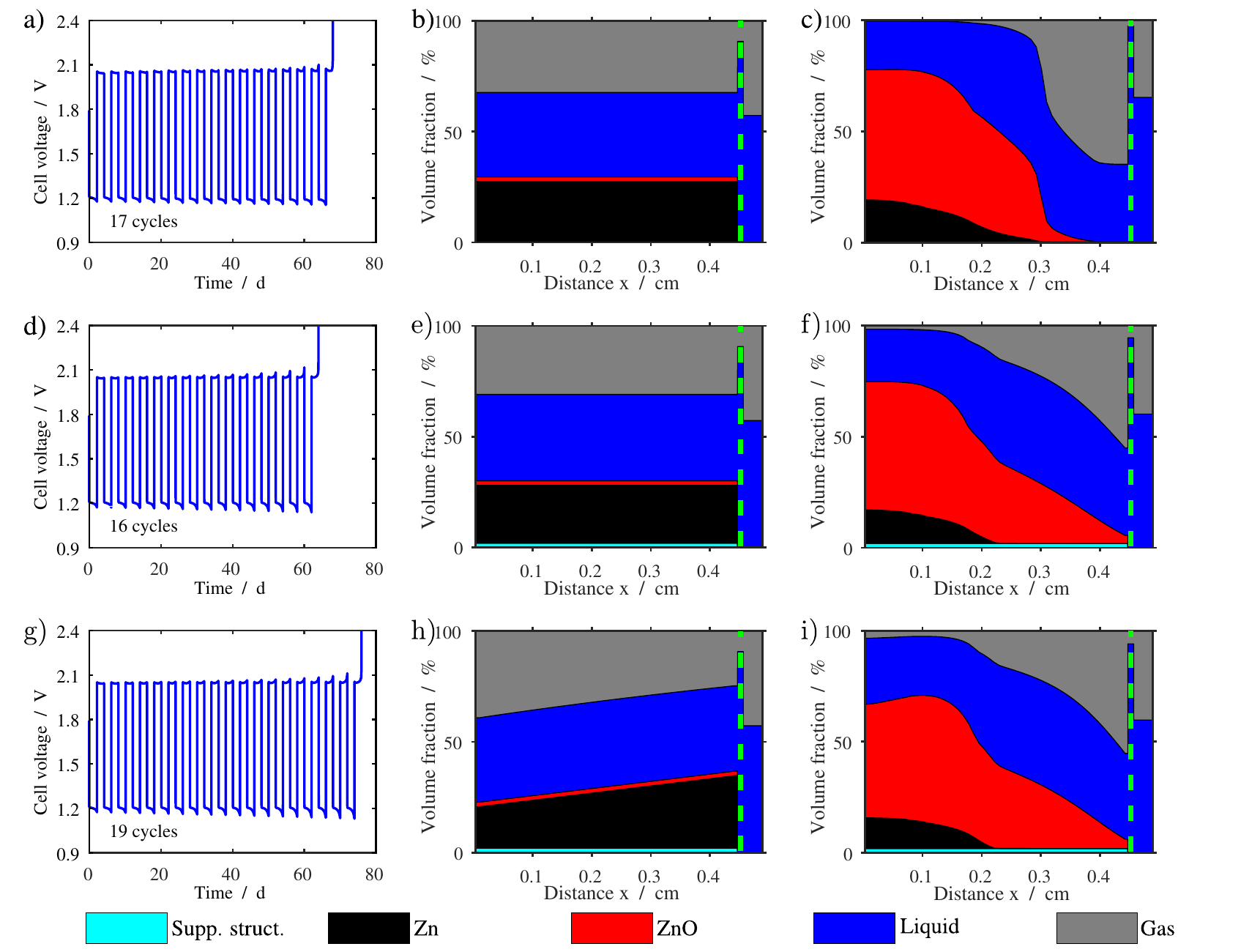}
    \caption[Optimization of initial configuration for advanced cyclability.]{Variation of initial configuration for advanced cycle-life. Each row represents an initial distribution of the solid phases depicted as radial projection in the central column and discussed in the main text. Initially, the volume fractions have a vanishing radial derivative ($\partial_{r} \varepsilon_{\alpha} = 0$). The first column shows the cell voltage over time and the last one the phase distribution in the cell at the last discharged state. Geometries (e) and (h) contain a supporting structure or 3D current collector, which provides a minimum specific surface area for reactions and suppresses solid convection. Cycling between $\SI{21}{\percent}$ and $\SI{90}{\percent}$ SOC (${T = \SI{298.15}{\kelvin}}$; ${\myvec{j} = \SI{100}{\ampere \per \meter\squared}}$).}
    \label{fig:ic:do:geometryTest}
\end{figure*}

Firstly, we assume a homogeneous initial $\zn$ distribution and achieve 17 cycles (see Figs. \ref{fig:ic:do:geometryTest}a-c). The solids cannot move freely without the gas reservoir below the lid which reduces cycle-life. In a second step, we add a supporting structure or 3D current collector in order to suppress solid convection (see Figs. \ref{fig:ic:do:geometryTest}d-i). To model this feature, we turn off solid convection (see \cref*{sec:suppl:params:levs}). For simplicity, we assume that this structure consists of inert $\zn$ material with a volume fraction of $\SI{1.92}{\percent}$. Our ideal supporting structure guarantees a minimum surface area for $\zn$ deposition and $\zn$ precipitation throughout the cell and an electron pathway in the solids through the electrode. Thus, we promote the more homogeneous formation of solid phases as shown in Figs. \ref{fig:ic:do:geometryTest}c,f,i. Nevertheless, cycle-life is even reduced slightly in this case. We note that our simple model for solid convection is agnostic of the particle nature of the $\zn$ electrode and tends to overestimate cycle-life. Future battery models could make use of hydrodynamic models for granular flow and include effects such as shear bands and anisotropic pressure distributions \cite{Latz2010_2}. In Figs. \ref{fig:ic:do:geometryTest}g-i, we use a $\zn$ distribution increasing linearly towards the cathode. The adjusted initial profile counteracts zinc shape-change and improves cycle-life up to 19 cycles. 

\subsection{Charge-Discharge Protocol Optimization}
\label{sec:ic:cdp}
\begin{figure}[!tb]
	\centering
	\includegraphics[width=0.95\columnwidth]{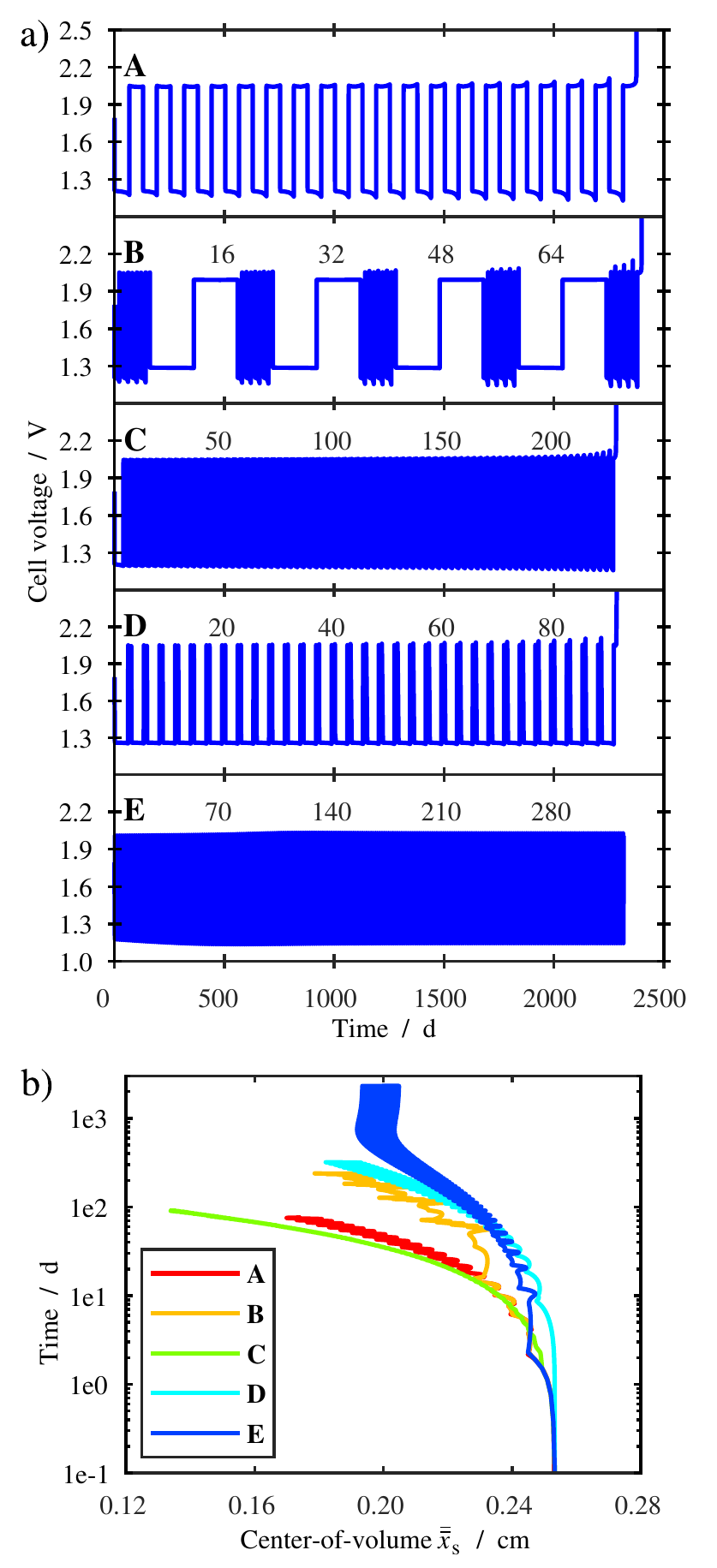}   
	\vspace*{-3mm}
	\caption[Improvement of the cyclability by different charge-discharge protocols.]{(a) Cell voltage showing improvement of cycle-life for the charge-discharge protocols in \cref{tab:ic:cdp:protocols}. (b) Effect of the charge-discharge protocol on the center-of-solid-volume movement (${T = \SI{298.15}{\kelvin}}$).}
	\label{fig:ic:cdp:cellVoltage_protocol}
\end{figure}

Smart charge-discharge protocols can improve cycle-life. Here, we discuss the effect of cycling range, applied charge/discharge currents, and recovery cycles. Table \ref{tab:ic:cdp:protocols} summarizes the applied protocols. The initial $\zn$ distribution for all simulations is depicted in \cref{fig:ic:do:geometryTest}h. As baseline reference, we use protocol \textbf{A}, whose simulated cycling behavior is shown in Figs. \ref{fig:ic:do:geometryTest}g-i. We show simulated cell voltages during cycling in \cref{fig:ic:cdp:cellVoltage_protocol}a and summarize the cycle-life in \cref{tab:ic:cdp:protocols}. The protocols \textbf{B}-\textbf{E} show a positive effect on the maximum number of cycles. However, not all extend the total utilizable capacity.

\begin{table}
	\centering
	\small
	\begin{tabular}{l | r r r r r}
		\toprule[1.5pt]		protocol number & \textbf{A} & \textbf{B} & \textbf{C} & \textbf{D} & \textbf{E} \\ 
		\midrule[2pt]
		$\myvec{j}_{\mathrm{charge}}$ / $\SI{}{\ampere \per\meter\squared}$ & 100 & 100 & 100 & 100 & 25 \\[4pt]		
		$\myvec{j}_{\mathrm{discharge}}$ / $\SI{}{\ampere \per\meter\squared}$ & 100 & 100 & 100 & 25 & 100 \\[4pt]
		min SOC / \% & 21 & 21 & 40 & 21 & 21 \\[4pt]
		max SOC / \% & 90 & 90 & 60 & 90 & 90 \\[4pt]
		recovery  & - & 5 & - & - & - \\[4pt] 
		$\myvec{j}^{\mathrm{recovery}}_{\mathrm{charge}}$ / $\SI{}{\ampere \per\meter\squared}$ & - & 10 & - & - & - \\[4pt]
		$\myvec{j}^{\mathrm{recovery}}_{\mathrm{discharge}}$ / $\SI{}{\ampere \per\meter\squared}$ & - & 10 & - & - & - \\
		\midrule[2pt]
		achieved cycles & 19 & 23 & 76 & 32 & $>$233 \\
		utilizable capacity / $\SI{}{\ampere \hour}$ & 3.9 & 3.9 & 4.5 & 6.5 & $>$48 \\
		\bottomrule[1.5pt]
	\end{tabular}
	\normalsize
	\caption[General setup of the charge-discharge protocols.]{General setup of the charge-discharge protocols with a galvanostatic operation mode. For the protocol \textbf{B} every 5-th cycle is a recovery cycle with $\myvec{j}^{\mathrm{recovery}}_{\mathrm{charge}}$ and ${\myvec{j}^{\mathrm{recovery}}_{\mathrm{discharge}}}$. The number of cycles in protocol \textbf{E} is only limited by a hard limit on the computation time, indicated by a ``$>$''.}
	\label{tab:ic:cdp:protocols}
\end{table}

Introducing recovery cycles \textbf{B}, in which the cell is charged and discharged very slowly, provides no advantage. This protocol increases the number of maximum cycles exactly by the number of recovery cycles yielding the same utilizable capacity. Since recovery cycles take extra time and waste energy they are even disadvantageous.

Cycling a smaller capacity range increases the number of cycles by a factor of four (see protocol \textbf{C}). Due to the smaller range, however, each discharge provides less capacity. In total, the utilizable capacity increases by $\SI{15}{\percent}$. Nevertheless, this protocol has a major drawback. It utilizes only $\SI{20}{\percent}$ of the total capacity and reduces the total energy density by a factor of four. 

Reducing the discharging current to $\SI{25}{\ampere \per \meter\squared}$ in \textbf{D} increases the utilizable capacity and the number of cycles by $\SI{66}{\percent}$. However, it is not always possible to reduce the discharging current, as it depends on the power specifications of the application, in which the battery is used. 

The most impressive protocol is \textbf{E}. By reducing the current density for charging, we increase the number of cycles and the total capacity by more than twenty-fold. This shows that cycling zinc-air batteries is limited by the charging current and that low charging currents are beneficial for cycle-life. 

Finally, we discuss the movement of the center-of-solid-volume for above protocols. \cref{fig:ic:cdp:cellVoltage_protocol}b shows the influence of the protocol on the movement of the solid phases. The most distinct shape-change effect is observed for the protocols \textbf{A} and \textbf{C}, whereas protocol \textbf{E} shows the smallest one. Very small capacity ranges are unfavorable as the inert electrolyte can prevent the back-shift of solids during a short charging period (see protocol \textbf{C}). Recovery cycles in protocol \textbf{B} indeed reduce the movement, but have no further special benefit. The best results are achieved for low charging currents as in protocol \textbf{E} due to the solid movement characteristics during charging and discharging. A large shift occurs during discharge, but the back-shift during charge compensates this shift for low charging currents. This demonstrates that the dissolution of $\zno$ is the critical process creating inhomogeneous $\zn$ deposition and irreversible agglomeration.

\subsection{Reaction and Transport Kinetics Optimization}
\label{sec:ic:ft}
Adjusting the intrinsic reaction and transport kinetics offers further options for improving cycling performance. Experimentally, this can for example be achieved by electrolyte additives, which promote $\zno$ dissolution, or additives to the $\zn$ paste, which promote homogeneous $\zn$ deposition. In the Supplementary Materials in \cref*{sec:suppl:ic}, we discuss this with two simulation approaches. Designing thicker anodes leads to a larger total $\zno$ surface area, improves $\zno$ dissolution, and avoids early $\znoh$ depletion. By doubling the anode thickness, we achieve $1.4$ times more cycles. Increasing the temperature to ${T = \SI{343.15}{\kelvin}}$ enhances the transport processes in the electrolyte. This makes it possible to achieve cycle numbers similar to protocols \textbf{D} and \textbf{E} without decreasing the charging current (see \cref{sec:ic:cdp}).

\section{Conclusion and Outlook}
\label{sec:mod:conc}
In this work, we present the first multi-dimensional model for conversion type batteries and apply it to zinc-air button cells. Local volume-averaging is the theoretical concept that enables us to model dissolution, growth, and movement of the porous zinc structure. Thus, we develop the most in-depth simulation study of inhomogeneous metal electrode structures so far.

Our model is successfully validated with x-ray tomography of commercial zinc-air cells. We demonstrate how electrolyte carbonation limits shelf-life and how zinc oxide nucleation limits rechargeability. The strong dependence of cycle-life on cell current is a major model prediction. In order to optimize cycling performance, we study various anode structures and examine the impact of charge-discharge protocols. Our simulations show that smaller cycling ranges and recovery cycles yield minor improvements. The best results are achieved for low current densities, especially low charging currents. Thus, dissolution of zinc oxide seems to be the critical process for cycle-life. As a consequence, accelerated intrinsic reaction and transport kinetics would enhance cycle-life as demonstrated by thicker anodes and higher operating temperature.

One-dimensional simulations, as by Stamm et al. \cite{Stamm2017}, remain relevant because the radial distribution of zinc and zinc oxide in coin cells has little influence on our simulation results. Nevertheless, it is well established in the field of lithium-ion batteries that multi-dimensional simulations are indispensable for studying degradation in inhomogeneous electrodes, for example, lithium plating \cite{Hein2016_1}. Thus, our novel simulation methodology paves the way for future insights into degradation of conversion batteries, for example, due to hydrogen evolution.

\subparagraph*{Conflicts of Interest}

There are no conflicts to declare.

\subparagraph*{Acknowledgments}

The authors thank Simon Hein and Max Schammer for fruitful discussions. This work was supported by the German Ministry of Education and Research (BMBF) (project LUZI, BMBF: 03SF0499E). Further support was provided by the bwHPC initiative and the bwHPCC5 project through associated compute services of the JUSTUS HPC facility at the University of Ulm. This work contributes to the research performed at CELEST (Center for Electrochemical Energy Storage Ulm-Karlsruhe).

\subparagraph*{References}

\bibliographystyle{elsarticle-num}

\bibliography{literature}

\end{document}